\documentclass[pra,twocolumn,tightenlines,nofootinbib]{revtex4}
\usepackage{graphicx}
\usepackage{multirow}
\usepackage{amsmath,amsfonts,amssymb}
\usepackage{bm}
\usepackage{dcolumn}
\usepackage{graphicx}

\newcommand{\eref}[1]{Eq.~(\ref{#1})}
\newcommand{\tref}[1]{Table~\ref{#1}}

\begin{document}

\title{Calculation of higher-order corrections to the light shift of the $5s^2\,\,^1\!S_0$ -\, $5s5p\,\,^3\!P_0^o$ clock transition in Cd}
\author{S.~G.~Porsev$^{1,2}$}
\author{M.~S.~Safronova$^{1,3}$}
\affiliation{$^1\!$Department of Physics and Astronomy, University of Delaware, Newark, Delaware 19716, USA\\
$^2\!$Petersburg Nuclear Physics Institute of NRC ``Kurchatov Institute'', Gatchina, Leningrad District, 188300, Russia\\
$^3\!$Joint Quantum Institute, NIST and the University of Maryland, College Park, Maryland 20742, USA
}
\date{\today}
\begin{abstract}
In the recent work [A.~Yamaguchi {\it et. al}, Phys. Rev. Lett. {\bf 123}, 113201 (2019)] Cd has been identified as an excellent candidate for a lattice clock. Here, we carried out computations needed for further clock development and made an assessment of the higher-order corrections to the light shift of the $5s^2\,\,^1\!S_0$ -\, $5s5p\,\,^3\!P_0^o$ clock transition. We carried out calculations of the magnetic dipole and electric quadrupole polarizabilities and linear and circular hyperpolarizabilities of the $5s^2\,\,^1\!S_0$ and $5s5p\,\,^3\!P_0^o$ clock states at the magic wavelength and estimated uncertainties of these quantities. We also evaluated the second-order Zeeman clock transition frequency shift.
\end{abstract}

\maketitle
\section{Introduction}
The Cd $5s^2\,\,^1\!S_0$ -\, $5s5p\,\,^3\!P_0^o$ transition has several desirable attributes for the development of a lattice clock. This clock has more than an order of magnitude smaller blackbody radiation (BBR) shift (a Stark shift resulting from the thermal radiation of the atoms environment, which is generally at 300 K temperature) in comparison with Sr and Yb~\cite{YamSafGib19,OvsMarPal16,DzuDer19}.
The size of a BBR shift is a property of the specific atomic transition used as a frequency standard and an uncertainty in the BBR shift is known to be one of the limiting systematic uncertainties in the clock uncertainty budget~\cite{NicCamHut15,HunSanLip16}. Short of cryogenic cooling, it cannot be suppressed and need to be quantified with high accuracy.

Two isotopes, $^{111}$Cd and $^{113}$Cd, both with 12\% natural abundance, have a nuclear spin of 1/2, which precludes tensor light shifts from the lattice light, another advantageous feature. Cd has the narrow $5s^2\,\,^1\!S_0$ -\, $5s5p\,\,^3\!P_1^o$ intercombination transition allowing Doppler cooling to 1.58 $\mu$K and simplifying a control of higher-order lattice light shifts~\cite{YamSafGib19}. The light for all of the transitions needed for the Cd clock, including the magic lattice, can be generated by direct, or frequency-doubled or quadrupled semiconductor lasers~\cite{YamSafGib19}.

In 2019, Cd clock magic wavelength was measured to be $419.88(14)$ nm \cite{YamSafGib19}, in excellent agreement with a  theoretical calculation reported in the same work. At magic wavelength, upper and lower clock states experience the same light shift, up to multipolar and higher-order effects considered in this work.  The fractional BBR shift was calculated to be $2.83(8) \times 10^{-16}$ at 300 K in Ref.~\cite{YamSafGib19}, in agreement with
Ref.~\cite{DzuDer19}. Recent progress opens a pathway to a rapid progress in Cd clock development and calls for a detailed investigation of the
clock systematic effects.

In this work we calculated properties needed to quantify higher-order light shifts:  magnetic dipole and electric quadrupole polarizabilities and linear and circular hyperpolarizabilities of the $5s^2\,\,^1\!S_0$ and $5s5p\,\,^3\!P_1^o$ clock states at the magic wavelength and estimated uncertainties of these quantities. We also evaluated the second-order Zeeman clock transition frequency shift in the presence of a weak magnetic field.

The paper is organized as follows. The general formalism and main formulas are presented in Section~\ref{Gen_for}. In Section~\ref{MetCalc}, we briefly describe the method of calculation. Section~\ref{results} is devoted to a discussion of the results obtained, and Section~\ref{concl} contains concluding remarks.
\section{General formalism}
\label{Gen_for}
We consider the Cd atom in a state $|0\rangle$ (with the total angular momentum $J=0$) placed in a field of the lattice standing wave with the electric-field vector given by
\begin{equation}
{\boldsymbol{\mathcal{E}}} = 2 {\boldsymbol{\mathcal{E}}_0}\, \mathrm{cos}(kx)\, \mathrm{cos}(\omega t) .
\label{field}
\end{equation}
Here $k=\omega/c$, $\omega$ is the lattice laser wave frequency, $c$ is the speed of light, and the factor 2 accounts for the superposition of forward and backward traveling along the x-axis waves. The atom-lattice interaction leads to the optical lattice potential for the atom that at $|kx| \ll 1$ can be approximated as~\cite{OvsMarPal16,PorSafSafKoz18}
\begin{eqnarray}
U(\omega) \approx &-&\alpha^{E1}(\omega)(1-k^2x^2)\,\mathcal{E}_0^2  \notag \\
                  &-&\{\alpha^{M1}(\omega) + \alpha^{E2}(\omega)\}k^2 x^2\,\mathcal{E}_0^2 \notag \\
                  &-&\beta(\omega)(1-2k^2x^2)\, \mathcal{E}_0^4 .
\label{DeltaE}
\end{eqnarray}
Here $\alpha^{E1}$, $\alpha^{M1}$, and $\alpha^{E2}$ are the electric dipole, magnetic dipole, and electric quadrupole polarizabilities, respectively, and $\beta$ is the hyperpolarizability defined below.

The ac $2^K$-pole polarizability of the $|0\rangle$ state with the energy $E_0$ is expressed
(we use atomic units $\hbar=m=|e|=1$) as~\cite{PorDerFor04}
\begin{eqnarray}
\alpha^{\lambda K}(\omega) &=& \frac{K+1}{K} \frac{2K+1}{[(2K+1)!!]^2}
(\alpha \,\omega)^{2K-2}  \notag \\
&\times& \sum_n \frac{(E_n-E_0) | \langle n||T_{\lambda K}||0 \rangle |^2}{(E_n-E_0)^2-\omega^2} ,
\label{Qlk}
\end{eqnarray}
where $\lambda$ stands for electric, $\lambda = E$, and magnetic, $\lambda =M$, multipoles and $\langle n||T_{\lambda K}||0 \rangle$ are
the reduced matrix elements of the multipole operators, $T_{E1} \equiv D$, $T_{M1} \equiv \mu$, and $T_{E2} \equiv Q$.

The expression for the hyperpolarizability of the $|0\rangle$ state depends on the polarization of the lattice wave. Below we consider the cases when the lattice wave is linearly or circularly polarized, and the 4th order correction to an atomic energy is determined by the
linear or circular hyperpolarizability, respectively.

The expression for the linear hyperpolarizability $\beta_l(\omega)$ is given by~\cite{PorSafSafKoz18}
\begin{equation}
\beta_l (\omega) = \frac{1}{9}\,Y_{101}(\omega) + \frac{2}{45}\,Y_{121}(\omega) ,
\label{beta_l}
\end{equation}%
with the quantities $Y_{101}\left( \omega \right) $ and $Y_{121}\left( \omega \right) $ determined as
\begin{eqnarray*}
\label{Y101}
Y_{101}(\omega) &\equiv& \sum_q \mathcal{R}_{101} (q\omega,2q\omega,q\omega) \notag \\
                     &+& \sum_{q,q'} \left[ \mathcal{R}'_{101}(q\omega,0,q'\omega)
                      - \mathcal{R}_1(q'\omega) \mathcal{R}_1(q\omega,q\omega) \right], \notag \\
Y_{121}(\omega) \! &\equiv& \! \sum_q \left[ \mathcal{R}_{121} (q\omega,2q\omega,q\omega)
                   + \sum_{q'} \mathcal{R}_{121}(q\omega,0,q'\omega) \right],
\end{eqnarray*}%
and $q,q'= \pm 1$.

The circular hyperpolarizability $\beta_c(\omega)$ can be written as
\begin{equation}
\beta_{c}=\frac{1}{9}\,X_{101}(\omega) + \frac{1}{18}\,X_{111}(\omega) + \frac{1}{15}\,X_{121}(\omega) ,
\label{beta_c}
\end{equation}
where
\begin{eqnarray*}
\label{X101}
X_{101}(\omega) &\equiv& \sum_{q,q'} \left[ \mathcal{R}'_{101}(q\omega,0,q'\omega)
                      - \mathcal{R}_1(q'\omega) \mathcal{R}_1(q\omega,q\omega) \right], \notag \\
X_{111}(\omega) &\equiv&  \sum_{q,q'} (-1)^{(q+q')/2}\, \mathcal{R}_{111}(q\omega,0,q'\omega), \notag \\
X_{121}(\omega) \! &\equiv& \! \sum_q \left[ \mathcal{R}_{121} (q\omega,2q\omega,q\omega)
                   + \frac{1}{6}\sum_{q'} \mathcal{R}_{121}(q\omega,0,q'\omega) \right] ,
\end{eqnarray*}%
and
\begin{widetext}
\begin{eqnarray}
\label{R_J1}
\mathcal{R}_{J_{m}J_{n}J_{k}}\left( \omega _{1},\omega _{2},\omega_{3}\right) &\equiv& \sum_{\gamma_m,\gamma_n,\gamma_k}
\frac{\left\langle \gamma _{0}J_{0}\left\Vert d\right\Vert \gamma_{m}J_{m}\right\rangle
\left\langle \gamma _{m}J_{m} \left\Vert d\right\Vert \gamma _{n}J_{n}\right\rangle
\left\langle \gamma _{n}J_{n}\left\Vert d\right\Vert \gamma _{k}J_{k}\right\rangle
\left\langle \gamma_{k}J_{k}\left\Vert d\right\Vert \gamma _{0}J_{0}\right\rangle }
{\left(E_{m}-E_{0}-\omega_1\right) \left( E_{n}-E_{0}-\omega_2\right) \left(E_k-E_0-\omega_3\right) }, \\
\label{R_J3}
\mathcal{R}_{J_m}(\omega) &\equiv& \sum_{\gamma_m} \frac{|\langle \gamma_0 J_0 ||d|| \gamma_m J_m\rangle|^2}{E_m-E_0-\omega},
\qquad
\mathcal{R}_{J_k} (\omega,\omega) \equiv \sum_{\gamma_k} \frac{|\langle \gamma_0 J_0 ||d|| \gamma_k J_k \rangle|^2}{(E_k-E_0-\omega)^2} .
\end{eqnarray}
\end{widetext}
The notation $\mathcal{R}^{\prime}_{101}$, i.e., the prime over $\mathcal{R}$, means that
the term $|\gamma_n\, 0 \rangle =|\gamma_0\, 0 \rangle$ (where $\gamma_n$ includes all other quantum numbers except $J$)
should be excluded from the summation over $\gamma_n$ in Eq.~(\ref{R_J1}).

The properties of the lattice potential for the Cd atom in its ground and excited clock states are determined
by \eref{DeltaE} and depend on the frequency. Below we analyze these properties at
the experimentally determined magic wavelength $\lambda^* = 419.88(14)$ nm~\cite{YamSafGib19}.
The magic frequency, $\omega^*$, corresponding to this wavelength, is $\omega^* \approx 23816$ cm$^{-1} \approx 0.108515 \,\,\mathrm{a.u.}$.

At the magic frequency the electric dipole polarizabilities of the clock $5s^2\, ^1\!S_0$ and $5s5p\, ^3\!P^o_0$ states are equal to each other, i.e.,
$\alpha^{E1}_{^1\!S_0}(\omega^*) = \alpha^{E1}_{^3\!P_0^o}(\omega^*)$. These polarizabilities were calculated in Ref.~\cite{YamSafGib19}
to be $63.7(1.9)$ a.u..

Using the formulas given above, we calculated the $M1$ and $E2$ polarizabilities and the linear and circular hyperpolarizabilities $\beta_{l,c}$ of the clock states at the magic frequency $\omega^*$, found respective differential polarizabilities and hyperpolarizabilities, and determined uncertainties of these values.
\section{Method of calculation}
\label{MetCalc}
We carried out calculations in the framework of high-accuracy relativistic methods combining configuration interaction (CI) with
(i) many-body perturbation theory (CI+MBPT method~\cite{DzuFlaKoz96}) and (ii) linearized coupled-cluster (CI+all-order method)~\cite{SafKozJoh09}.
In these methods the energies and wave functions are found from the multiparticle Schr\"odinger equation
\begin{equation}
H_{\mathrm{eff}}(E_n) \Phi_n = E_n \Phi_n,  \label{Heff}
\end{equation}
where the effective Hamiltonian is defined as
\begin{equation}
H_{\mathrm{eff}}(E) = H_{\mathrm{FC}} + \Sigma(E).
\end{equation}
Here, $H_{\mathrm{FC}}$ is the Hamiltonian in the frozen core approximation and $\Sigma$ is the energy-dependent correction, which takes into account virtual core excitations in the second order of the perturbation theory (the CI+MBPT method) or in all orders of the perturbation theory
(the CI+all-order method).

To accurately calculate the valence parts of the polarizabilities and hyperpolarizabilities, we solve the inhomogeneous equation using the Sternheimer~\cite{Ste50} or Dalgarno-Lewis~\cite{DalLew55} method following formalism developed in Ref.~\cite{KozPor99}. We use an effective (or ``dressed'') electric-dipole operator in our calculations that includes the random-phase approximation (RPA). To calculate such complicated quantities as $\mathcal{R}_{J_m J_n J_k}$ and carry out accurately three summations over intermediate states, we solve the inhomogeneous equation twice. A detailed description of this approach is given in Ref.~\cite{PorSafSafKoz18}.
\section{Results and discussion}
\label{results}
We carried out calculations of the $M1$ and $E2$ polarizabilities and the hyperpolarizabilities in the CI+MBPT and CI+all-order approximations. In both cases the theoretical energies were used. The CI+all-order calculations include higher-order terms in comparison with the CI+MBPT calculations and are more accurate. The difference of these two calculations gives us an estimate of the uncertainty of the results.
\subsection{Linear and circular hyperpolarizabilities of the $^1\!S_0$ and $^3\!P_0^o$ clock states}
In calculating quantities given by Eqs.~(\ref{R_J1}) and (\ref{R_J3}) a main contribution comes from valence electrons. The core electrons contribution is much smaller and we included it only to $\mathcal{R}_1(\omega)$ terms.

Indeed, as follows from Eq.~(\ref{R_J3}), the quantity $\mathcal{R}_1 (\omega,\omega)$
can be treated as the derivative of $\mathcal{R}_1(\omega)$ over $\omega$, i.e.,
\begin{equation*}
\mathcal{R}_1 (\omega,\omega) = \frac{\partial \mathcal{R}_1 (\omega)}{%
\partial \omega} = \underset{\Delta \rightarrow 0}{\lim} \frac{\mathcal{R}%
_1(\omega +\Delta) - \mathcal{R}_1(\omega)}{\Delta}.
\end{equation*}
Since the core contribution to $\mathcal{R}_{1}(\omega )$ is rather insensitive to $\omega $ and $\Delta $ is small, the core contributions to $\mathcal{R}_{1}(\omega +\Delta )$ and $\mathcal{R}_{1}(\omega )$ are practically identical and cancel each other in the expression for $\mathcal{R}_{1}(\omega, \omega )$.

Taking into account the uncertainty of our results for the $^{1}\!S_{0}$ and $^{3}\!P_{0}^{o}$ hyperpolarizabilities, we assume that the core contribution to the $\mathcal{R}_{1Jn1}(\omega _{1},\omega _{2},\omega_{3})$ terms is also negligible. This assumption is based on the calculation of the static hyperpolarizability for the Sr$^{2+}$ ground state that was found to be 62.6 a.u.~\cite{YuSuoFen15}. This is negligibly small compared to valence contribution to $\mathcal{R}_{1Jn1}(\omega _{1},\omega _{2},\omega_{3})$
in case of the quite similar $5s^2\,^1\!S_0$ and $5s5p\,^3\!P_0^o$ clock states in Sr~\cite{PorSafSafKoz18}.

The results of calculation of the linear and circular hyperpolarizabilities of the $^1\!S_0$ and $^3\!P_0^o$ clock states
are presented in Table~\ref{Tab:beta}.
\begin{table}[tbp]
\caption{Contributions to the linear and circular hyperpolarizabilities $\beta_{l,c} (5s^2\,^1\!S_0)$ and $\beta_{l,c}(5s5p\,^3\!P_0^o)$ (in a.u.) calculated in the CI+all-order (labeled as ``CI+All'') and CI+MBPT (labeled as ``CI+PT'') approximations at the magic frequency
$\omega^* = 0.108515$ a.u.. The ``Total'' values are obtained according to Eqs.~(\ref{beta_l}) and (\ref{beta_c}).
$\Delta \beta_{l,c} \equiv \beta_{l,c}(^3\!P^o_0) - \beta_{l,c}(^1\!S_0)$ is the difference of the ``Total'' $^3\!P^o_0$ and $^1\!S_0$ values.
Numbers in brackets represent powers of 10. The uncertainties are given in parentheses.}
\label{Tab:beta}%
\begin{ruledtabular}
\begin{tabular}{llrrrr}
 && \multicolumn{2}{c}{$5s^2\,^1\!S_0$} & \multicolumn{2}{c}{$5s5p\,^3\!P_0^o$} \\
 &Contrib. &\multicolumn{1}{c}{CI+All} &\multicolumn{1}{c}{CI+PT} &\multicolumn{1}{c}{CI+All} &\multicolumn{1}{c}{CI+PT} \\
\hline \\ [-0.3pc]
$\beta_l$ & $\frac{1}{9}Y_{101}(\omega)$   & 3.61[4] & 2.71[4] & -5.30[5] & -5.37[5]  \\[0.4pc]
          & $\frac{2}{45}Y_{121}(\omega)$  & 5.64[4] & 5.08[4] &  4.37[5] &  4.81[5]  \\[0.4pc]
          & Total                          & 9.24[4] & 7.80[4] & -9.23[4] & -5.61[4]  \\[0.3pc]
$\Delta \beta_l$                           && \multicolumn{1}{c}{-1.85[5]} & \multicolumn{1}{c}{-1.34[5]} & \\
                                           & Recommended & \multicolumn{2}{c}{$-1.85(50) \times 10^5$}   && \\[0.3pc]
                                           & Ref.~\cite{OvsMarPal16} &\multicolumn{2}{c}{$-10.2 \times 10^5$} & & \\[0.3pc]
\hline \\ [-0.6pc]
$\beta_c$ & $\frac{1}{9}X_{101}(\omega)$   &-1.98[4] &-1.88[4] & -6.03[5]  & -5.95[5]  \\[0.4pc]
          & $\frac{1}{18}X_{111}(\omega)$  &   41    &   34    &  7.21[6]  &  6.61[6]  \\[0.4pc]
          & $\frac{1}{15}X_{121}(\omega)$  & 6.21[4] & 5.53[4] & -1.45[6]  & -1.11[6]  \\[0.4pc]
          & Total                          & 4.23[4] & 3.66[4] &  5.15[6] &  4.90[6]  \\[0.3pc]
$\Delta \beta_c$                           &&  5.11[6] &  4.86[6] &        &          \\[0.3pc]
                                           & Recommended &\multicolumn{2}{c}{$5.11(25) \times 10^6$} & & \\[0.3pc]
                                           & Ref.~\cite{OvsMarPal16} &\multicolumn{2}{c}{$3.65 \times 10^6$} & &
\end{tabular}
\end{ruledtabular}
\end{table}
Our recommended value of differential linear hyperpolarizability,
$\Delta \beta_l (\omega^*)= -1.85(50) \times 10^{5} \,\, {\rm a.u.}$, is two orders of magnitude smaller (in absolute value) than analogous differential hyperpolarizability for Sr, $\Delta \beta_l = -1.5(4)\times 10^{7}$ a.u.~\cite{PorSafSafKoz18}. In case of Cd, the absolute values of the contributing terms are generally smaller than in Sr, and there are significant cancellations between them.

The circular hyperpolarizability of the $^3\!P_0^o$ state is two orders of magnitude larger in absolute value than the circular hyperpolarizability of the $^1\!S_0$ state and the linear hyperpolarizability of the $^3\!P_0^o$ state. This is explained as follows: the main contribution to $\beta_c(^3\!P_0^o)(\omega^*)$ comes from the term
\begin{widetext}
\begin{equation*}
\mathcal{R}_{111}(\omega^*,0,\omega^*)  \equiv \sum_{\gamma_m,\gamma_n,\gamma_k}
\frac{\langle ^3\!P_0^o ||d|| \gamma_m J_m=1 \rangle \langle \gamma_m J_m=1 ||d|| \gamma_n J_n=1 \rangle
\langle \gamma_n J_n=1 ||d|| \gamma_k J_k=1 \rangle \langle \gamma_k J_k=1 ||d|| ^3\!P_0^o \rangle}
{(E_m - E_{^3\!P_0^o} -\omega^*) (E_n - E_{^3\!P_0^o}) (E_k - E_{^3\!P_0^o} - \omega^*)} .
\end{equation*}
\end{widetext}
In the sum over $\gamma_n$ there is the intermediate state $5s5p\,\,^3\!P_1^o$ separated from $^3\!P_0^o$ by the fine-structure interval. In this case the energy denominator $E_{^3\!P_1^o} - E_{^3\!P_0^o} \approx 542\,\,{\rm cm}^{-1}$ is small and, respectively, the contribution of this term is large, leading to much larger hyperpolarizability for the circular polarization.

We compare our results with those obtained in Ref.~\cite{OvsMarPal16} in \tref{Tab:beta}. There is a reasonable agreement
for differential circular hyperpolarizability while our differential linear hyperpolarizability is 5 times smaller in absolute value
than that found in Ref.~\cite{OvsMarPal16}.
\begin{table}[t]
\caption{The dynamic $M1$ and $E2$ polarizabilities (in a.u.) of the $5s^2\,^1\!S_0$ and $5s5p\,^3\!P_0^o$ states at the magic frequency,
calculated in the CI+MBPT (labeled as ``CI+MBPT'') and CI+all-order (labeled as ``CI+All'') approximations. The recommended
value of $\Delta \alpha^{QM}$ is given in the line ``Recom. $\Delta \alpha^{QM}$''. The uncertainties are given in parentheses.}
\label{Tab:alpha}%
\begin{ruledtabular}
\begin{tabular}{ccc}
Polariz.                  &    CI+MBPT            &      CI+All               \\
\hline \\ [-0.5pc]
$\alpha^{M1}(^1\!S_0)$    &  1.5$\times 10^{-9}$  &  1.6$\times 10^{-9}$      \\[0.3pc]
$\alpha^{M1}(^3\!P^o_0)$  & -4.0$\times 10^{-6}$  & -3.9$\times 10^{-6}$      \\[0.3pc]
$\Delta \alpha^{M1}$      & -5.5$\times 10^{-6}$  & -5.5$\times 10^{-6}$       \\[0.5pc]

$\alpha^{E2}(^1\!S_0)$    &  2.29$\times 10^{-5}$ &  2.43(14)$\times 10^{-5}$   \\[0.3pc]
$\alpha^{E2}(^3\!P^o_0)$  &  8.97$\times 10^{-5}$ &  8.88(8)$\times 10^{-5}$   \\[0.3pc]
$\Delta \alpha^{E2}$      &  6.68$\times 10^{-5}$ &  6.45(23)$\times 10^{-5}$    \\[0.5pc]

$\Delta \alpha^{QM}$      &  6.28$\times 10^{-5}$ &  6.05(23)$\times 10^{-5}$   \\[0.7pc]
Recom. $\Delta \alpha^{QM}$ &                     &  6.05(23)$\times 10^{-5}$  \\[0.5pc]
Ref.~\cite{OvsMarPal16}   &                       &  3.13 $\times 10^{-5}$
\end{tabular}
\end{ruledtabular}
\end{table}
\subsection{$M1$ and $E2$ polarizabilities at the magic frequency}
\label{M1andE2}
To accurately calculate the valence part of the $E2$ polarizabilities of the clock states at the magic frequency, we solved inhomogeneous equation with the electric quadrupole operator $Q$ in the right hand side. As in the case of hyperpolarizability, we calculated these quantities using both the CI+all-order and CI+MBPT methods, including the RPA corrections to the operator $Q$. The core contributions were calculated in the RPA. For the $M1$ polarizabilities, only a few low-lying intermediate states give dominant contributions, and it is sufficient to calculate their sum. We estimate the uncertainties of the results as the difference between the CI+all-order and CI+MBPT values.

The final values of the polarizabilities and their uncertainties are listed in Table~\ref{Tab:alpha}. We also determined the recommended value of $\Delta \alpha^{QM} \equiv \Delta \alpha^{E2} + \Delta \alpha^{M1}$, where
\begin{eqnarray}
\Delta \alpha^{M1} &\equiv& \alpha^{M1}(^3\!P^o_0) - \alpha^{M1}(^1\!S_0),  \notag \\
\Delta \alpha^{E2} &\equiv& \alpha^{E2}(^3\!P^o_0) - \alpha^{E2}(^1\!S_0) .
\label{delEM}
\end{eqnarray}

To determine the uncertainty of $\Delta \alpha^{QM}$ we note that the $\alpha^{M1}(^1\!S_0)$ polarizability is very small and we can neglect it. The $\alpha^{M1}(^3\!P^o_0)$  polarizability is more than three orders of magnitude larger in absolute value than $\alpha^{M1}(^1\!S_0)$, but still an order of magnitude smaller than $\Delta \alpha^{E2}$. Therefore, the uncertainty of $\Delta \alpha^{QM}$ is mostly determined by the uncertainty in $\Delta \alpha^{E2}$ and we estimate it to be 4\%. Comparing our recommended value for $\Delta \alpha^{QM}$ with the result obtained in Ref.~\cite{OvsMarPal16}, we see that there is a fair agreement between them.
\subsection{Second order Zeeman shift}
In this section we consider a systematic effect due to second order Zeeman shift which both clock states
experience in the presence of a weak external magnetic field. If an atom is placed in a such magnetic field $\mathbf{B}$,
the interaction of the atomic magnetic moment $\bm \mu$ with $\mathbf{B}$ is described by the Hamiltonian
\begin{equation}
H = - {\boldsymbol \mu} \cdot \mathbf{B}.
\label{eq:magnetic_H}
\end{equation}
The atomic magnetic moment $\bm \mu$ is mostly determined by the electronic magnetic moment and can be written as
\begin{equation}
 \bm \mu = -\mu_0 ({\bf J} + {\bf S}),
\end{equation}
where ${\bf J}$ and ${\bf S}$ are the total and spin angular momenta of the atomic state and $\mu_0$ is the Bohr magneton
defined as $\mu_0 = |e| \hbar/(2mc)$.

Directing the external magnetic field $\mathbf{B}$ along the $z$-axis (${\bf B} = B_z \equiv B$),
we calculate the second order Zeeman shift, $\Delta E$, (in the absence of hyperfine interaction) as
\begin{equation}
\Delta E = -\frac{1}{2} \alpha^{\rm M1} B^2,
\label{DelE}
\end{equation}
where $\alpha^{\rm M1}$ is the magnetic-dipole polarizability. For a state $|J=0\rangle$ it is reduced to the scalar polarizability, given by
\begin{equation}
\alpha^{\rm M1} = \frac{2}{3} \sum_n \frac{|\langle n || \mu || J=0 \rangle|^2}{E_n-E_0}.
\label{alphaM1}
\end{equation}

To estimate the second order Zeeman shift for the clock transition
$$\Delta \nu \equiv \frac{\Delta E(^3\!P^o_0) - \Delta E(^1\!S_0)}{h}$$
we note that the $\alpha^{M1}(^1\!S_0)$  polarizability is negligibly small compared to $\alpha^{M1}(^3\!P^o_0)$,
so we can write $\Delta \nu \approx \Delta E(^3\!P^o_0)/h$.

For an estimate of $\alpha^{M1}(^3\!P^o_0)$ we take into account that the main contribution to this polarizability comes from the intermediate
state $5s5p\,\,^3\!P_1^o$. Then, from \eref{alphaM1} we obtain
\begin{equation}
\alpha^{\rm M1}(^3\!P^o_0) \approx \frac{2}{3} \frac{\langle ^3\!P^o_1 || \mu || ^3\!P^o_0 \rangle^2}{E_{^3\!P^o_1}-E_{^3\!P^o_0}} .
\end{equation}
Using for an estimate
\begin{equation}
|\langle ^3\!P^o_1 || \mu || ^3\!P^o_0 \rangle| \approx \sqrt{2} \mu_0
\end{equation}
and substituting it to \eref{DelE} we find
\begin{equation}
\Delta E (^3\!P^o_0) \approx -\frac{2}{3} \frac{\mu_0^2}{E_{^3\!P^o_1}-E_{^3\!P^o_0}}\,B^2
\label{DE3P0}
\end{equation}
in agreement with the result obtained in Ref.\cite{BoyZelLud07}.

Using the experimental value of energy difference $E_{^3\!P^o_1}-E_{^3\!P^o_0} \approx 542\,\, {\rm cm}^{-1}$, we arrive at
$$\Delta \nu \approx -80\, B^2,$$ where $\Delta \nu$ is in mHz and the magnetic field $B$ is in G.
\section{Conclusion}
\label{concl}
We carried out calculations of the magnetic dipole and electric quadrupole polarizabilities as well as linear and circular hyperpolarizabilities of the clock
$5s^2\,\,^1\!S_0$ and $5s5p\,\,^3\!P_0^o$ states at the magic wavelength and compared them with other available data. We also evaluated the second-order Zeeman shift for the clock transition frequency.
These values are required for an assessment of the higher-order corrections to the light shift of the $5s^2\,\,^1\!S_0$ -\, $5s5p\,\,^3\!P_0^o$ clock transition.
We have demonstrated that the linear differential hyperpolarizability for the clock transition for Cd is two orders of magnitude smaller than for Sr and Yb. We also found the circular hyperpolarizability to be much larger than the linear hyperpolarizability and explained the source of this difference. A knowledge of the multipolar polarizabilities and hyperpolarizabilities at different polarizations of the lattice wave is needed for further Cd clock development and selection of the lattice configurations to minimize the higher-order light shifts.

\section{Acknowledgements}
We thank Kurt Gibble for helpful discussions. This work was supported by the Office of Naval Research under Grant No. N00014-17-1-2252.
S.G.P acknowledges support by the Russian Science Foundation under Grant No.~19-12-00157.


\end{document}